\documentclass[a4paper,11pt]{article}
\usepackage{jheppub} 
\usepackage{lineno}
\usepackage{amsfonts}
\usepackage{amsmath}
\usepackage{amssymb}
\usepackage{amsthm}
\usepackage{bbm}

\def\[#1\]{\begin{align}#1\end{align}}

\def \nn {\nonumber}

\def \dd{\mathrm{d}}

\title{\boldmath Timelike Entanglement Entropy in Higher Curvature Gravity}

\author{Zi-Xuan Zhao$^{a}$ 
,
Long Zhao$^{e,f,a,}$\footnote{Corresponding author}
, Song He$^{b,c,d,}$\footnote{Corresponding author}}
\affiliation{$^{a}$Center for Theoretical Physics and College of Physics, Jilin University,\\ Changchun 130012, People's Republic of China\\
$^{b}$Institute of Fundamental Physics and Quantum Technology,
Ningbo University, \\Ningbo, 315211, People's Republic of China\\
$^{c}$School of Physical Science and Technology, Ningbo University,\\ Ningbo, 315211, People's Republic of China\\
$^{d}$Max Planck Institute for Gravitational Physics (Albert Einstein Institute),\\
Am M\"uhlenberg 1, 14476 Golm, Germany\\
$^{e}$Research Center for Quantum Physics and Technologies, Inner Mongolia University, \\
Hohhot 010021, People's Republic of China\\
$^{f}$School of Physical Science and Technology, Inner Mongolia University, \\Hohhot 010021, People's Republic of China\\}

\emailAdd{zzx23@mails.jlu.edu.cn;  zhaolong@jlu.edu.cn; hesong@nbu.edu.cn}

\abstract{This work investigates holographic timelike entanglement entropy in higher curvature gravity, with a particular focus on Lovelock theories and on the role of excited states. For strip subsystems, higher-curvature terms are found to affect the imaginary part of the entropy in a dimension-dependent manner, while excited states contribute solely to the real part. For the cases analyzed, spacelike and timelike entanglement entropies exhibit proportional relations: vacuum contributions differ by universal phase factors, while excitation contributions are linked by dimension–dependent rational coefficients. For hyperbolic subsystems, the timelike entanglement entropy computed via complex extremal surfaces is shown to agree with results obtained through analytic continuation, with imaginary contributions appearing in all dimensions. Higher-curvature corrections are explicitly calculated in five- and $(d+1)$-dimensional Gauss-Bonnet gravity, illustrating the applicability of the complex surface prescription to general Lovelock corrections. These results provide a controlled setting to examine the influence of higher-curvature interactions on holographic timelike entanglement entropy, and clarify its relation to vacuum and excited-state contributions.}

\begin{document}
\maketitle
\flushbottom

\section{Introduction}
The AdS/CFT correspondence~\cite{Maldacena:1997re, Gubser:1998bc, Witten:1998qj} (also known as gauge/gravity duality) provides a powerful framework linking gravitational theories in an asymptotically anti-de Sitter space to a conformal field theory on its boundary. Its discovery has motivated much research related to quantum information theory in the high-energy physics community in recent years. Among them, entanglement entropy, as a carrier of quantum information, has emerged as a pivotal concept in modern theoretical physics, acting as a bridge between quantum information theory and gravitational dynamics~\cite{Casini:2004bw, Calabrese:2004eu,Kitaev:2005dm,Casini:2016fgb,Nishioka:2018khk,Witten:2018zxz,Casini:2022rlv}. In the context of the AdS/CFT correspondence, the celebrated Ryu-Takayanagi formula elucidates how the entanglement entropy of a spatial region in a boundary conformal field theory corresponds to the area of an extremal (minimal) surface in the bulk spacetime~\cite{Ryu:2006bv, Ryu:2006ef,Hubeny:2007xt}. This geometric realization underscores that spacetime might emerge from quantum entanglement patterns~\cite{VanRaamsdonk:2010pw}.

Recently, the notion of timelike entanglement entropy—where the boundary subregion extends along a timelike instead of spacelike direction—has been introduced~\cite{Doi:2022iyj,Doi:2023zaf}. It naturally takes a complex-valued form and can be interpreted as a form of pseudo-entropy~\cite{Nakata:2020luh,Mollabashi:2021xsd,Doi:2022iyj,Parzygnat:2023avh,Narayan:2023ebn}, generalizing conventional entanglement measures. The relationship between timelike entanglement entropy, pseudo-entropy, and spacelike entanglement entropy in the context of dS/CFT has been discussed in \cite{Narayan:2022afv,Jiang:2023loq,Narayan:2023zen,Nanda:2025tid}. Beyond merely serving as an analytic continuation of spacelike entanglement entropy, timelike entanglement entropy has been assigned a physical interpretation in~\cite{Milekhin:2025ycm}: it corresponds to the pseudo-entropy of the transition matrix between two spacelike subsystems separated by a timelike interval. In the literature~\cite{Guo:2024lrr,Xu:2024yvf,Guo:2025mwp,Guo:2025pru}, the authors found that the timelike entanglement entropy for a timelike subregion $t\in[0,t_0]$ can be expressed as the spacelike entanglement entropy for a spacelike subregion $x\in[-t_0,t_0]$. In a black hole background, the extremal surface for a timelike subsystem crosses the event horizon, while the extremal surface for a spacelike subsystem remains outside the horizon. This relation implies that information inside the horizon can be probed solely using information from outside the horizon. For other recent advances in this field, see~\cite{Das:2023yyl,Basu:2024bal,Anegawa:2024kdj,Afrasiar:2024ldn,He:2024emd,Wen:2024yny,vjyt-xc15,Roychowdhury:2025ebs,Roychowdhury:2025ukl,Katoch:2025bnh,Chu:2025sjv,Ikeda:2025gju,Nunez:2025ppd}.

In 3-dimensional holography, \cite{Doi:2023zaf} proposed that partly spacelike and partly timelike bulk geodesics whose respective real and imaginary lengths reproduce the analytic continuation of the entanglement entropy of a single subregion. Since timelike entanglement entropy in quantum field theory can be defined by an analytic continuation~\cite{Doi:2023zaf}, it should come as no surprise that holographically the relevant geometric notion will be an analytic continuation of the extremal surfaces geometrizing entanglement entropy, such that they are anchored on a timelike subregion. In~\cite{Heller:2024whi}, the authors identified that such extremal surfaces will be in general complex, i.e., they perceive the bulk geometry for complex rather than real spacetime coordinates. In other words, timelike entanglement is captured by complex extremal surfaces extending into analytically continued (complex) bulk geometries, offering a novel temporal probe into the fabric of spacetime.

Realistic quantum gravity scenarios often entail higher-curvature corrections. These corrections significantly modify the holographic entanglement entropy formula—for example, replacing the area functional with generalized Wald-like entropy expressions that include extrinsic curvature contributions~\cite{Wald:1993nt, Iyer:1995kg, Jacobson:1993vj}. {As in Reference~\cite{Sun:2008uf}, the author first discussed the modification of holographic entanglement entropy due to the Chern-Simons term.} However, \cite{Hung:2011xb} shows that in general Wald's formula for horizon entropy does not yield the correct entanglement entropy. Fortunately, for Lovelock gravity, there is an alternative prescription \cite{Jacobson:1993xs} that involves only the intrinsic curvature of the bulk surface and has been proven to correctly reproduce the universal contribution to entanglement for CFTs in 4 and 6 dimensions. For arbitrary higher-derivative gravity theories, the authors, following the approach of~\cite{Lewkowycz:2013nqa}, derived the holographic entanglement entropy formula by computing the semi-classical gravitational path integral~\cite{Dong:2013qoa,Dong:2016hjy}. This offers a potential avenue for investigating the timelike entanglement entropy in higher-curvature gravity. {For arbitrary higher-derivative gravity theories, the authors, following the approach of~\cite{Lewkowycz:2013nqa}, derived the holographic entanglement entropy formula by computing the semi-classical gravitational path integral~\cite{Dong:2013qoa}.}

As a preliminary exploration, this work would investigate timelike entanglement entropy within the framework of Lovelock gravity \cite{Lovelock:1970zsf, Lovelock:1971yv}. Lovelock gravity represents the most general extension of Einstein gravity in higher dimensions that preserves second-order field equations, making it a natural theoretical laboratory for exploring quantum gravity effects beyond the Einstein–Hilbert action. 
Since higher-curvature terms generically arise as low-energy corrections in string theory and other ultraviolet completions of gravity, understanding timelike entanglement entropy in such theories provides a more realistic holographic description of entanglement phenomena in quantum gravity\footnote{We remark that quadratic Lovelock gravity (Gauss--Bonnet) does arise naturally in the 
low-energy effective action of heterotic string theory, where $\mathcal{O}(\alpha')$ 
curvature-squared corrections can be recast into the Gauss--Bonnet density by 
consistency with ghost-freeness \cite{Zwiebach:1985uq}. By contrast, 
higher-derivative corrections in string theory, such as the well-known $R^4$ terms 
in type IIB \cite{Kiritsis:1997em} or analogous combinations in other compactifications, 
do not generally take the form of pure Lovelock densities.
}. Moreover, timelike entanglement entropy itself extends the concept of spatial entanglement entropy to timelike-separated regions, yielding complex-valued entanglement measures that probe the temporal structure of correlations. Studying timelike entanglement entropy in Lovelock gravity\footnote{One important reason for focusing on Lovelock gravity is that the form of holographic entanglement entropy in this theory has already been
systematically developed, which provides us with a convenient framework for
extending the study to other quantum information measures. We acknowledge,
however, that the method in this paper has limitations when directly applied to
more general higher–derivative gravities.}, therefore, offers a unique opportunity to understand how higher-curvature interactions modify the geometry of complex extremal surfaces and affect the real and imaginary parts of holographic entanglement measures.

The structure of the paper is as follows. Section~\ref{section2} provides a brief review of the relevant background. Section~\ref{section3} presents the analysis of timelike entanglement entropy for a strip-like subsystem in Lovelock gravity. In Section~\ref{section4}, the study is extended to include hyperbolic subsystems. Section~\ref{section5} concludes with a summary of results and further discussion.

\section{A few preliminaries}\label{section2}
The primary objective of this work is to examine the effects of higher-curvature interactions in the bulk gravitational theory on holographic timelike entanglement entropy. The analysis is carried out within the framework of Lovelock gravity~\cite{Lovelock:1971yv, Lovelock:1970zsf}, which provides a tractable model for explicit computations. To provide the necessary background, this section includes a brief review of timelike entanglement entropy and the relevant aspects of Lovelock gravity.

\subsection{Timelike entanglement entropy }
 Timelike entanglement entropy provides a natural extension of the standard entanglement entropy to timelike-separated subsystems, offering new insights into the causal structure of quantum correlations. It is defined by analytically continuing the entanglement entropy to a timelike subsystem $A$, denoted $S_A^{(T)}$. In two-dimensional quantum field theory in a flat spacetime, for a spacelike interval $A$ with endpoints $A_1 = (t_1,x_1)$ and $A_2 = (t_2,x_2)$, the entanglement entropy is 
 \[
 S_A=\frac{c}{3}\log\left[\frac{\sqrt{(x_1-x_2)^2-(t_1-t_2)^2}}{\epsilon}\right],\label{equ2.1.1}
 \]
 where $\epsilon$ is a UV regulator. Analytically continuing \eqref{equ2.1.1} to the timelike case $(x_1-x_2)^2-(t_1-t_2)^2 < 0$, yields
\[
 S_A^{(T)}=\frac{c}{3}\log\left[\frac{\sqrt{-(x_1-x_2)^2+(t_1-t_2)^2}}{\epsilon}\right]+\frac{c\pi }{6}i.
 \]
In particular, for a purely timelike interval, i.e., $x_1-x_2=0$ and $t_1-t_2=\Delta t$, one finds
 \[
 S_A^{(T)}=\frac{c}{3}\log\left[\frac{\Delta t}{\epsilon}\right]+\frac{c\pi }{6}i.\label{equ2.1.3}
 \]
In three-dimensional holography, \cite{Doi:2023zaf, Doi:2022iyj} proposed a geometric interpretation in which the real part of \eqref{equ2.1.3} is reproduced by the length of a spacelike geodesic, while the length of a timelike geodesic reproduces the imaginary part. The holographic timelike entanglement entropy is then given by
\[
S_A^{(T)}=\frac{\mathrm{Area}(\gamma_{A})}{4G},\label{equ2.1.4}
\]
where $G$ is the bulk gravitational constant. 
This prescription reproduces the $AdS_3$ result \eqref{equ2.1.3} and extends naturally to higher dimensions. Further discussions and generalizations can be found in \cite{Bhattacharya:2012mi, Nunez:2025ppd, Heller:2025kvp, Gong:2025pnu}.

Based on this prescription, the analysis proceeds within the framework of Lovelock gravity, which offers a tractable higher-curvature extension for studying holographic timelike entanglement entropy beyond Einstein gravity.

\subsection{Lovelock gravity}
Lovelock gravity \cite{Lovelock:1970zsf, Lovelock:1971yv} is a higher-dimensional generalization of Einstein’s theory that incorporates higher-curvature interactions proportional to the Euler densities of even-dimensional manifolds. The general Lovelock action in $d+1$ dimensions is given by
\[
I=\frac{1}{2\ell_P^{d-1}}\int \dd^{d+1}x \sqrt{-g}\left[\frac{d(d-1)}{L^2}+R+\sum\limits_{p=2}^{\lfloor\frac{d+1}{2}\rfloor}c_p L^{2p-2} \mathcal{L}_{2p}(R)\right],\label{equ2.2.1}
\]
where $\lfloor\frac{d+1}{2}\rfloor$ denotes the integer part of $(d+1)/2$ and $c_p$ are dimensionless coupling constants for the higher curvature terms $\mathcal{L}_{2p}(R)$. These higher-order interactions are defined as
\[
\mathcal{L}_{2p}(R)\equiv \frac{1}{2^p}\delta_{\mu_1\mu_2\dots\mu_{2p-1}\mu_{2p}}^{\nu_1\nu_2\dots\nu_{2p-1}\nu_{2p}} R^{\mu_1\mu_2}_{\quad\quad \nu_1\nu_2}\dots R^{\mu_{2p-1}\mu_{2p}}_{\quad\quad\quad \nu_{2p-1}\nu_{2p}},\label{L2P}
\]
which is proportional to the Euler density on a $2p$-dimensional manifold. Here, the symbol $\delta_{\mu_1\mu_2\dots\mu_{2p}}^{\nu_1\nu_2\dots\nu_{2p}}$ is used to denote the totally antisymmetric product of $2p$ Kronecker delta symbols. The cosmological constant and the Einstein terms can be incorporated into the scheme as $\mathcal{L}_0$ and $\mathcal{L}_1$, respectively. However, the explicit expressions are provided above to establish the normalization of both the Planck length and the length scale $L$. By construction, it is clear that in $d+1$ dimensions, all Lovelock $\mathcal{L}_p$ terms with $p>(d+1)/2$ must vanish, hence the explicit restriction on the sum in eq. \eqref{equ2.2.1} is not really required. For $p=(d+1)/2$,  $\mathcal{L}_p$ is topological and does not contribute to the gravitational equations of motion.

In anticipation of applications to the AdS/CFT correspondence, a negative cosmological constant is explicitly included in the action \eqref{equ2.2.1}. The theory then admits $AdS_{d+1}$ vacua with curvature scale $\Tilde{L}^2=L^2/f_{\infty}$ where $f_{\infty}$ is a root of:
\[
1=f_\infty-\sum\limits_{p=2}^{\lfloor\frac{d+1}{2}\rfloor}\lambda_p(f_\infty)^p,\label{equ2.2.2}
\]
and the coefficients $\lambda_p$ are defined as
\[
\lambda_p=(-)^p\frac{(d-2)!}{(d-2p)!}c_p.
\]
Equation~\eqref{equ2.2.2} generally admits $\lfloor d/2 \rfloor$ distinct roots for $f_\infty$. The analysis is restricted to positive real roots, which correspond to $AdS_{d+1}$ vacua. In the regime of small $\lambda_p$ couplings, the relevant solution is the smallest positive root, continuously connected to the Einstein gravity value $f_\infty = 1$ in the limit $\lambda_p \to 0$. To ensure a smooth connection with the Einstein gravity limit while capturing higher-derivative gravitational corrections to timelike entanglement entropy, the discussion is confined to this small-coupling regime and focuses exclusively on the corresponding root.

\section{Timelike entanglement entropy for a strip-like subsystem in Lovelock gravity}\label{section3}
This section investigates timelike entanglement entropy in Lovelock gravity. The analysis begins with five–dimensional Gauss–Bonnet gravity, which offers a tractable setting for computing leading–order corrections. It is then extended to arbitrary dimensions to reveal universal patterns in Gauss–Bonnet modifications. The discussion proceeds to seven–dimensional Lovelock gravity—the minimal case admitting cubic curvature interactions—before considering finite–order Lovelock truncations in general $(d+1)$–dimensional spacetimes. Finally, higher–curvature corrections in the timelike case are compared with their spacelike counterparts.

The strip subsystem of interest lies in $d$-dimensional Minkowski spacetime, located on the regulated$(z=\epsilon\ll1)$ boundary of the bulk metric
\[
ds^2=\frac{\Tilde{L}^2}{z^2}\left(-f(z)dt^2+\frac{dz^2}{f(z)}+d\mathbf{x}^2\right).
\]
The choice $f(z)=1$ corresponds to the empty AdS space, which describes the vacuum of the dual CFT. The strip is defined by
\[
A=\left\{(t,\mathbf{x}): t\in \left[-\frac{\Delta t}{2},\frac{\Delta t}{2}\right], \mathbf{x}_{\|}\in\mathbb{R}^{d-2}, x_\perp=0\right\}.
\]
For $d>2$, the holographic timelike entanglement entropy in the vacuum is known \cite{Doi:2023zaf}:
\[
S_A^{(T)}=\frac{\left(\frac{1}{\epsilon^{d-2}}+\frac{c_d}{2}\frac{(-i)^d}{(\Delta t)^{d-2}}\right)}{2(d-2)G},\quad c_d=\left(\frac{2\sqrt{\pi}\Gamma\left(\frac{d}{2(d-1)}\right)}{\Gamma\left(\frac{1}{2(d-1)}\right)}\right)^{d-1},\label{TEE}
\]
and the corresponding codimension-two bulk surface $\gamma_{A}$ takes the form\cite{Heller:2024whi}
\[
\mathbf{X}^\mu=\left\{t_{\pm}(z),z, \mathbf{x}_{\|}, x_\perp=0\right\},
\]
with
\[
t_\pm(z)=&A_\pm \pm i\frac{z_t}{d}\left(\frac{z}{z_t}\right)^d\times {}_2F_1\left(\frac{1}{2},\frac{d}{2(d-1)},\frac{3d-2}{2(d-1)},\left(\frac{z}{z_t}\right)^{2d-2}\right)\nn\\
&A_\pm=\pm\frac{\Delta t}{2}, \quad z_t=\frac{i \Gamma\left(\frac{1}{2(d-1)}\right)}{2\sqrt{\pi}\Gamma\left(\frac{d}{2(d-1)}\right)} \Delta t.\label{RTsurface}
\]
\subsection{Timelike entanglement entropy in five-dimensional Gauss-Bonnet gravity}
Five-dimensional Lovelock gravity—also known as Gauss–Bonnet gravity—can be obtained by adding the Gauss–Bonnet term to the Einstein–Hilbert action. The theory is described by
\[
I=\frac{1}{2 \ell_p^3}\int \dd^5x\sqrt{-g}\left[R+\frac{12}{L^2}+\frac{\lambda_5 L^2}{2} L_4\right]
\]
where 
\[
L_4=R_{\mu\nu\rho\sigma}R^{\mu\nu\rho\sigma}-4R_{\mu\nu}R^{\mu\nu}+R^2\label{L4}
\]
is the Gauss–Bonnet density. Here, $\lambda_5$ is the Gauss–Bonnet coupling, and $L$ denotes the curvature radius of the AdS background. In AdS Gauss–Bonnet gravity, the theory admits a pure AdS solution \cite{Boulware:1985wk, Cai:2001dz, Hung:2011xb},
\[
ds^2=\frac{\Tilde{L}^2}{z^2}\left(-dt^2+dz^2+dx_1^2+dx_2^2+dx_3^2\right)\label{equ3.1.1}
\]
where $\Tilde{L}^2$ is the effective AdS radius, related to $L $ by 
\[
\Tilde{L}^2=L^2/f_\infty,\quad f_\infty=\frac{1-\sqrt{1-4\lambda_5}}{2\lambda_5}.
\]
The holographic entanglement entropy formula for Gauss-Bonnet gravity has been discussed in \cite{deBoer:2011wk, Hung:2011xb}, which can be expressed as
\[
S_A=\frac{2\pi}{\ell_p^3}\int_M\dd^3x\sqrt{h}\left[1+\lambda_5L^2\mathcal{R}\right]+\frac{4\pi}{\ell_p^3}\int_{\partial M}\dd^2 x \sqrt{\gamma}\lambda_5 L^2 \mathcal{K},\label{equ3.1.2}
\]
where the first integral is evaluated on the extremal surface $M$, the second one is on $\partial M$, which is the boundary of $M$ regularized at $z=\epsilon$. $\mathcal{R}$ is the Ricci scalar for the intrinsic geometry of $M$, and $\mathcal{K}$ is the trace of the extrinsic curvature of $\partial M$. $h$ is the determinant of the induced metric on $M$ while $\gamma$ is the determinant of the induced metric on $\partial M$. The ``Gibbons-Hawking'' boundary term is added in eq. \eqref{equ3.1.2} to ensure a well-defined variational principle in extremizing the functional. 

At present, there is no general formula for holographic timelike entanglement entropy in higher-derivative gravity. However, inspired by the proposal of \cite{Heller:2024whi}, timelike entanglement entropy can be associated with the area of a complex extremal surface. This observation provides the basis for extending the Ryu–Takayanagi prescription—originally defined for real, spacelike extremal surfaces—to the complex extremal surfaces appearing in eq.~\eqref{equ3.1.2}, thereby offering a holographic interpretation of timelike entanglement entropy in Gauss–Bonnet gravity.


Beyond the pure AdS geometries of eq.~\eqref{equ3.1.1}, timelike entanglement entropy may also be investigated for excited states in conformal field theories. Following the logic of \cite{Bhattacharya:2012mi}, the gravity dual of such an excited state can be described by
\[
ds^2=\frac{\Tilde{L}^2}{z^2}\left(-f(z)dt^2+\frac{dz^2}{f(z)}+dx_1^2+dx_2^2+dx_3^2\right),\label{equ3.1.3}
\]
with $f(z)\approx 1-mz^4$, where $m$ characterizes the near-boundary deviation from the pure AdS metric\footnote{We recall that eq.~\eqref{equ3.1.3} corresponds to the Gauss--Bonnet--AdS black brane solution \cite{Cai:2001dz}, with 
\[
f(z)=\frac{1}{2\lambda_5}\left(1-\sqrt{1-4\lambda_5\left(1-\frac{z^4}{z_h^4}\right)}\right),
\]
where $z_h$ denotes the horizon of the black brane. Near the AdS boundary the metric 
function behaves as $f(z)\approx 1 - m z^4$ with $m = 1/z_h^4$, which is the form 
employed in our analysis. Importantly, our argument does not rely on assumptions 
about the infrared region $z\to\infty$ \cite{Bhattacharya:2012mi}. Since the 
Gauss--Bonnet--AdS black brane corresponds holographically to a thermal state of 
the boundary CFT, the parameter $m$ measures the energy density of the excitation. 
In particular, the small-$m$ regime can be interpreted as a low-thermal excitation, 
i.e.\ a slightly thermalized state close to the vacuum. Our analysis in this paper is therefore focused on the small-$m$ regime, where 
a controlled perturbative expansion of timelike entanglement entropy can be performed, 
while the study of generic $m$ corresponding to fully thermalized states is left for 
future work.
}. Unless stated otherwise, all higher-curvature gravitational corrections in this work refer to timelike entanglement entropy in such excited states. $m$ reflects the asymptotic behavior of the gravity background near the boundary. {It is challenging to obtain an exact expression for the timelike entanglement entropy in a black hole background within Lovelock gravity. To make progress, both the excitation parameter $m$ and the Lovelock couplings $c_p$ are treated as small perturbative parameters. In this regime, the method of~\cite{Guo:2013aca} can be adopted, expanding the entropy as a series in these small quantities:
\[
S_A(\mathcal{M},\alpha)
=&S_A(\mathcal{M}_0,0)+\left.\frac{\delta S_A(\mathcal{M}_0,\lambda)}{\delta\lambda_i}\right|_{\lambda=0}\lambda_i
+\left.\frac{\delta^2S(\mathcal{M}_0,\lambda)}{\delta\lambda_i\delta\lambda_j}\right|_{\lambda=0}\lambda_i\lambda_j\nn\\
&{
+\left.\frac{\delta^2S(\mathcal{M},\lambda)}{\delta\mathcal{M}\delta\lambda_i}\right|_{\mathcal{M}_0,\lambda=0}\frac{\delta\mathcal{M}}{\delta\lambda_j}\lambda_i\lambda_j
+\left.\frac{\delta^2S(\mathcal{M},0)}{\delta\mathcal{M}^2}\right|_{\mathcal{M}_0}\frac{\delta\mathcal{M}}{\delta\lambda_i}\frac{\delta\mathcal{M}}{\delta\lambda_j}\lambda_i\lambda_j}+\cdots\,,\label{variation}
\]
where $\lambda$ collects all parameters ($m$, $c_p$), $\mathcal{M}$ represents the exact solution of the extremal surface, and $\mathcal{M}_0$ represents the solution of the extremal surface when $m, c_p=0$. Since $m$ and $c_p$ are independent parameters, and we wish to simultaneously consider the effects of both higher-derivative gravitational corrections and the excitation, we retain terms in the above expression up to and including order $O(m)O(c_p)$, while discarding terms of order $O(m^2)$ or $O(c_p^2)$.
}

The induced metric on the complexified bulk surface is
\[
ds^2_{strip}=\frac{\Tilde{L}^2}{z^2}\left(\left(1+mz^4-\left(1-mz^4\right)\Dot{t}^2\right)dz^2+dx_1^2+dx_2^2\right),
\]
where a dot denotes the derivative with respect to $z$. Carrying out the computation, the holographic timelike entanglement entropy for the excited state \eqref{equ3.1.3} in Gauss–Bonnet gravity, following eq.~\eqref{equ3.1.2}, can be expressed as
\[
S_A^{(T)}=\frac{2\pi\Tilde{L}^3}{\ell_p^3}\int_\epsilon^{z_t'}\dd z\frac{(1+mz^4+2f_\infty\lambda_5+(-1+mz^4)\Dot{t}^2(z))}{z^3\sqrt{(1+mz^4+(-1+mz^4)\Dot{t}^2(z))}},\label{equ3.1.4}
\]
where the volume along $x_1$ and $x_2$ is normalized to unity {and $z_t'$ is the maximal value of $z$ on the surface in the bulk which is controlled by 
\[
\Delta t=\int_\epsilon^{z_t'}\Dot{t}\dd z
\]
with minimizing the functional \eqref{equ3.1.4} whose e.o.m. is
\[
\frac{\Dot{t} \left(m z^4-1\right) \left(-2 f_\infty\lambda_5 +m \left(\Dot{t} ^2+1\right) z^4-\Dot{t} ^2+1\right)}{z^3 \left(m \left(\Dot{t} ^2+1\right) z^4-\Dot{t} ^2+1\right)^{3/2}}=-\frac{1}{z_t'^3}.\label{eom1}
\]
}
Although $\lambda_5$ is not required to vanish, the construction guarantees that the vacuum result \eqref{TEE} is exactly recovered in the limit $\lambda_5 \to 0$, consistent with holographic duality.

{The e.o.m. \eqref{eom1} admits a solution 
\[
\Dot{t}=(1+2f_\infty\lambda_5)\left(\frac{z^6}{z^6-z_t'^6}\right)^\frac{1}{2}+\frac{m \left(2 z^6-3 z_t'^6\right) \left(\frac{z^6}{z^6-z_t'^6}\right)^{3/2}}{2 z^2}
\]
when $f_\infty \lambda_5$ and $m$ are treated as a small parameters.
}

For the excited state, $S_A^{(T)}$ can be expanded as a double series~\eqref{variation} in $\lambda_5$ and $m$ around the point $(0,0)$. By substituting $t(z)$ from the complex extremal surface~\eqref{RTsurface}, the leading-order gravitational corrections in Gauss–Bonnet gravity are obtained
\[
S_A^{(T)}
=&\frac{\left(\frac{1}{\epsilon^{2}}+\frac{c_4}{2}\frac{1}{(\Delta t)^{2}}\right)}{4G}
+\frac{f_\infty\lambda_5}{4G}\left(\frac{2}{\epsilon^2}-\frac{4 \pi ^{3/2} \Gamma \left(\frac{2}{3}\right)^3}{\Delta t^2\Gamma \left(\frac{1}{6}\right)^2 \Gamma \left(\frac{7}{6}\right)}\right)\nn\\
-&\frac{\Delta t^2 m \Gamma \left(\frac{1}{6}\right)^2 \Gamma \left(\frac{4}{3}\right)}{8G \left(\sqrt{\pi } \Gamma \left(\frac{2}{3}\right)^2 \Gamma \left(\frac{5}{6}\right)\right)}
+\frac{15 \sqrt{\pi } f_\infty \lambda_5 m \Delta t^2 \Gamma \left(\frac{7}{6}\right) \Gamma \left(\frac{7}{3}\right)}{8 G \Gamma \left(\frac{2}{3}\right)^2 \Gamma \left(\frac{5}{6}\right) \Gamma \left(\frac{11}{6}\right)}+\dots\label{result3.1}
\]
where $``\dots"$ represents the subleading contribution in Gauss-Bonnet gravity and $c_4$ is defined in \eqref{RTsurface}. In eq.~\eqref{result3.1}, the first term reproduces the vacuum holographic timelike entanglement entropy \eqref{TEE} without gravitational corrections. The second and fourth terms capture higher-curvature corrections, while the third and fourth terms encode contributions from low excited states. The Gauss-Bonnet coupling $\lambda_5$ simultaneously enhances the UV area-law coefficient and modifies the coefficient of the $(\Delta t)^{-2}$ ``finite'' geometric term, reflecting a universal reweighting of the vacuum contribution by higher-curvature effects. In contrast, low-energy excitations enter only at order $\Delta t^2$, and their impact is further modulated by $\lambda_5$. This indicates that higher-curvature corrections can either amplify or suppress the timelike entanglement entropy response to excitations, depending on the physically allowed range of $\lambda_5$.


\subsection{Timelike entanglement entropy in $d+1$-dimensional Gauss-Bonnet gravity}\label{subsection3.2}
After analyzing the five-dimensional Gauss–Bonnet case as a reference, the discussion is extended to Gauss–Bonnet gravity in arbitrary $(d+1)$ dimensions to examine the corresponding corrections to timelike entanglement entropy. In this context, the Lovelock series~\eqref{equ2.2.1} truncates at $p_{\text{max}} = 2$, and the action takes the form
\[
I=\frac{1}{2\ell_P^{d-1}}\int \dd^{d+1}x \sqrt{-g}\left[\frac{d(d-1)}{L^2}+R+\frac{L^2\lambda}{(d-2)(d-3)}\mathcal{L}_4\right].
\]
The $AdS_{d+1}$ metric
\[
ds^2=\frac{\Tilde{L}^2}{z^2}\left(-dt^2+dz^2+\sum\limits_{i=1}^{d-1}+dx_i^2\right)
\]
is an exact solution to the equations of motion.

The holographic timelike entanglement entropy functional for Gauss-Bonnet gravity is
\[
S_A^{(T)}=\frac{2\pi}{\ell_p^{d-1}}\int_M\dd^{d-1}x\sqrt{h}\left[1+\frac{2L^2\lambda}{(d-2)(d-3)}\mathcal{R}\right]+\frac{4\pi}{\ell_p^{d-1}}\int_{\partial M}\dd^{d-2} x \sqrt{\gamma}\frac{2L^2\lambda}{(d-2)(d-3)}\mathcal{K},\label{equ3.2.1}
\]
where $M$ is the complexified extremal surface and $\partial M$ its regulated boundary at $z=\epsilon$. For excited states, the dual gravity background can be modeled as
\[
ds^2=\frac{\Tilde{L}^2}{z^2}\left(-f(z)dt^2+\frac{dz^2}{f(z)}+\sum\limits_{i=1}^{d-1}+dx_i^2\right),
\]
with $f(z)\approx 1-mz^d$. The induced metric on the complexified bulk surface is then
\[
ds^2_{strip}=\frac{\Tilde{L}^2}{z^2}\left(\left(1+mz^d-\left(1-mz^d\right)\Dot{t}^2\right)dz^2+\sum\limits_{i=1}^{d-2}+dx_i^2\right).
\]
Using the standard warped–geometry formulas \cite{Misner:1973prb}, the intrinsic curvature and extrinsic curvature of the surface are given by
\[
&\mathcal{R}=-\frac{(d-2)\left[(d-1)+(2d-1)mz^d+\left(-(d-1)+(2d-1)mz^d\right)\Dot{t}^2+2z\left(-1+mz^d\right)\Dot{t}\Ddot{{t}}\right]}{\Tilde{L}^2\left(1+mz^d+\left(-1+mz^d\right)\Dot{t}^2\right)^2}\nn\\
&\mathcal{K}=\frac{d-2}{\Tilde{L}}\sqrt{\frac{1}{1+mz^d+(-1+mz^d)\Dot{t}^2}}.
\]
Substituting these into \eqref{equ3.2.1} yields
\[
S_A^{(T)}=\int_\epsilon^{z_t'}\dd z\frac{2 \pi  \Tilde{L}^{d-1} \left(2f_\infty \lambda +\left(m z^{d}-1\right)\Dot{t}^2+m z^{d}+1\right)}{\ell_p^{d-1} z^{d-1} \sqrt{\left(m z^{d}-1\right) \Dot{t}^2+m z^{d}+1}}\label{equ3.2.2}
\]
where the volume of $\mathbb{R}^{d-2}$ spanned by $x_1\dots x_{d-2}$ is normalized to unity. { Again, the e.o.m. derived from the functional \eqref{equ3.2.2} 
\[
\frac{\Dot{t} z^{1-d} \left(m z^d-1\right) \left(m \left(\Dot{t}^2+1\right) z^d-2 f_\infty\lambda -\Dot{t}^2+1\right)}{\left(m \left(\Dot{t}^2+1\right) z^d-\Dot{t}^2+1\right)^{3/2}}=-\frac{1}{z_t'^{d-1}}
\]admits a perturbative solution of the form
\[
\Dot{t}=(1+2f_\infty\lambda)\left(\frac{z^{2d-2}}{z^{2d-2}-z_t'^{2d-2}}\right)^\frac{1}{2}+\frac{m \left(2 z^{2d-2}-3 z_t'^{2d-2}\right) \left(\frac{z^{2d-2}}{z^{2d-2}-z_t'^{2d-2}}\right)^{3/2}}{2 z^{d-2}}
\]when $f_\infty \lambda$ and $m$ are treated as a small parameters.}

By expanding $S_A^{(T)}$ in the parameters $\lambda$ and $m$ as in \eqref{variation} about $(0,0)$, and inserting $t(z)$ from the complexified extremal surfaces \eqref{RTsurface}, the leading–order gravitational corrections to timelike entanglement entropy in $(d+1)$–dimensional Gauss–Bonnet gravity are obtained:
\[
S_A^{(T)}=&\frac{\left(\frac{1}{\epsilon^{d-2}}+\frac{c_d}{2}\frac{(-i)^d}{(\Delta t)^{d-2}}\right)}{2(d-2)G}+\frac{f_\infty\lambda}{(d-2)^2G}\left(\frac{2}{ \epsilon^{d-2}}-\frac{i^{2-d} c_d}{2 \Delta t^{d-2}}\frac{\Gamma\left(\frac{d}{2(d-1)}\right)^2}{\Gamma\left(\frac{1}{2(d-1)}+1\right)\Gamma\left(\frac{1}{2(d-1)}\right)}\right)\nn\\
+&\frac{m \Delta t^{2}}{8(d-2)G}\left(\frac{1}{d+1}-\frac{2f_\infty\lambda}{d-3}\right)\frac{\Gamma\left(\frac{1}{2(d-1)}\right)^2\Gamma\left(\frac{1}{(d-1)}\right)}{\sqrt{\pi}\Gamma\left(\frac{1}{(d-1)}-\frac{1}{2}\right)\Gamma\left(\frac{d}{2(d-1)}\right)^2}+\dots\label{result3.2}
\]
where $``\dots"$ represents the subleading contribution in Gauss-Bonnet gravity. 

This result provides the timelike entanglement entropy in arbitrary spacetime dimension $d$, organized into three distinct contributions. The first line contains the vacuum divergences: the standard UV divergence $\epsilon^{-(d-2)}$ together with a $(\Delta t)^{-(d-2)}$ term that originates from the analytic continuation of the spacelike expression. The Gauss-Bonnet (or more generally higher-curvature) coupling $\lambda$, encoded through $f_\infty$, universally rescales both of these contributions, in particular shifting the coefficient of the area-law term. The second line represents the leading correction due to low-energy excitations of mass parameter $m$, scaling as $\Delta t^2$; higher-curvature effects again modulate this term through the factor $\big(1/(d+1) - 2 f_\infty \lambda/(d-3)\big)$. The appearance of dimension-dependent Gamma-function ratios reflects the nontrivial continuation from spacelike to timelike intervals, ensuring that both the divergent and finite parts respect the expected analytic structure across arbitrary dimensions. Importantly, in the expression for~\eqref{result3.2}, gravitational corrections in $(d+1)$-dimensional Gauss-Bonnet gravity may acquire an imaginary part\footnote{The imaginary part of timelike entanglement entropy is closely connected to the notion of pseudo-entropy in de Sitter space \cite{Doi:2022iyj}, and the higher-curvature corrections obtained here may be interpreted as higher-derivative gravitational modifications of this quantity. 
}, whereas the excited state contributions remain strictly real-valued, contributing only to the real part of the entropy.

\subsection{Timelike entanglement entropy in seven-dimensional Lovelock gravity}
The case of a six-dimensional boundary is now considered. In this setting, the bulk spacetime is seven-dimensional, and both curvature-squared and curvature-cubed terms contribute to the Lovelock action~\eqref{equ2.2.1}, leading to
\[
I=\frac{1}{2\ell_p^5}\int\dd^7 x \sqrt{-g}\left[\frac{30}{L^2}+R+\frac{L^2}{12}\lambda_7 \mathcal{L}_4-\frac{L^4}{24}\mu_7 \mathcal{L}_6\right],
\]
where $\mathcal{L}_4$ is given in \eqref{L4}, and $\mathcal{L}_6$ can be evaluated as
\[
\mathcal{L}_6=&4R_{\mu\nu}^{\quad\rho\sigma}R_{\rho\sigma}^{\quad\tau\chi}R_{\tau\chi}^{\quad\mu\nu}-8R_{\mu\ \ \nu}^{\ \ \rho\  \ \sigma}R_{\rho\ \ \sigma}^{\ \ \tau\  \ \chi}R_{\tau\ \ \chi}^{\ \ \mu\  \ \nu}-24R_{\mu\nu\rho\sigma}R^{\mu\nu\rho}_{\ \quad\tau}R^{\sigma\tau}+3R_{\mu\nu\rho\sigma}R^{\mu\nu\rho\sigma}R\nn\\
&+24R_{\mu\nu\rho\sigma}R^{\mu\rho}R^{\nu\sigma}+16R_{\mu}^{\ \nu}R_{\nu}^{\ \rho}R_{\rho}^{\ \mu}-12R_{\mu}^{\ \nu}R_{\nu}^{\ \mu}R+R^3
\]
using \eqref{L2P}. 
Seven-dimensional Lovelock gravity admits a pure AdS solution with effective radius $\Tilde{L}^2=L^2/f_\infty$, where $f_\infty$ is the smallest positive root of
\[
1=f_\infty-f_\infty^2\lambda_7-f_\infty^3\mu_7.
\]

The holographic timelike entanglement entropy in seven-dimensional Lovelock gravity is a natural extension of the higher-curvature holographic entanglement entropy formula discussed in \cite{Hung:2011xb} and takes the form
\[
S_A^{(T)}=\frac{2\pi}{\ell_p^{5}}\int_M\dd^{5}x\sqrt{h}\left[1+\frac{\lambda_7L^2}{6}\mathcal{R}-\frac{\mu_7 L^4}{8}\left(\mathcal{R}_{\mu\nu\rho\sigma}\mathcal{R}^{\mu\nu\rho\sigma}-4\mathcal{R}_{\mu\nu}\mathcal{R}^{\mu\nu}+\mathcal{R}^2\right)\right]+\text{surfaceterm}\label{equ3.3.1}
\]
where $h$ is the determinant of the induced metric on the complexified bulk surface $M$. Following \cite{Myers:1987yn}, the surface term is
\[
\text{surfaceterm}=\frac{2\pi}{\ell_p^{5}}\int_{\partial M}\dd^{4}x\sqrt{\gamma}\left[\frac{\lambda_7L^2}{3}\mathcal{K}-\frac{\mu_7 L^4}{8}\left(4\mathcal{R}^B\mathcal{K}-8\mathcal{R}_{ij}^B\mathcal{K}^{ij}-\frac{4}{3}\mathcal{K}^3+4\mathcal{K}\mathcal{K}_{ij}\mathcal{K}^{ij}-\frac{8}{3}\mathcal{K}_{ij}\mathcal{K}^{jk}\mathcal{K}_k^i\right)\right],
\]
where $\partial M$ is the boundary of $M$, $\gamma$ is the determinant of the induced metric on $\partial M$, $\mathcal{K}_{ij}$ and $\mathcal{K}$ are the extrinsic curvature and its trace on boundary $\partial M$, $\mathcal{R}_{ij}^B$ and $\mathcal{R}^B$ are the intrinsic Ricci tensor and Ricci scalar of the boundary $\partial M$ respectively.

An excitation of pure AdS in seven-dimensional Lovelock gravity is considered, described by
\[
ds^2=\frac{\Tilde{L}^2}{z^2}\left(-f(z)dt^2+\frac{dz^2}{f(z)}+dx_1^2+dx_2^2+dx_3^2+dx_4^2+dx_5^2\right)
\] 
with $f(z)\approx1-mz^6$. Its induced metric on $M$ is then
\[
ds^2_{strip}=\frac{\Tilde{L}^2}{z^2}\left(\left(1+mz^6-\left(1-mz^6\right)\Dot{t}^2\right)dz^2+dx_1^2+dx_2^2+dx_3^2+dx_4^2\right).
\]
A direct computation yields the timelike entanglement entropy functional for this low-excitation state:
\[
&S_A^{(T)}=\frac{2\pi\Tilde{L}^5}{\ell_p^5}\int_\epsilon^{z_t'}\dd z\frac{1}{ z^5 \left(\left(m z^6-1\right) \Dot{t}^2+m z^6+1\right)^{3/2}}\nn\\
&\left(f_\infty^2 \mu_7+2 \left(m z^6-1\right) \Dot{t}^2 \left(f_\infty \lambda_7+m z^6+1\right)+\left(m z^6+1\right) \left(2 f_\infty \lambda_7+m z^6+1\right)+\left(m z^6-1\right)^2 \Dot{t}^4\right),\label{equ3.3.2}
\]
where the volume of $\mathbb{R}^4$  is normalized to unity. 
{The e.o.m. derived from the functional \eqref{equ3.3.2} 
\[
\frac{\Dot{t} \left(m z^6-1\right) \left(\left(m \left(\Dot{t}^2+1\right) z^6-\Dot{t}^2+1\right) \left(-2 f_\infty \lambda_7 +\Dot{t}^2 \left(m z^6-1\right)+m z^6+1\right)-3 f_\infty^2 \mu_7 \right)}{z^5 \left(m \left(\Dot{t}^2+1\right) z^6-\Dot{t}^2+1\right)^{5/2}}=-\frac{1}{z_t'^{5}}
\]admits a perturbative solution of the form
\[
\Dot{t}=&(1+2f_\infty\lambda_7)\left(\frac{z^{10}}{z^{10}-z_t'^{10}}\right)^\frac{1}{2}+\frac{m \left(2 z^{10}-3 z_t'^{10}\right) \left(\frac{z^{10}}{z^{10}-z_t'^{10}}\right)^{3/2}}{2 z^{4}}\nn\\
-&3f_\infty^2\mu_7\frac{z^{10}}{z_t'^{10}}\left(\frac{z^{10}}{z^{10}-z_t'^{10}}\right)^{-\frac{1}{2}}
\]when $f_\infty \lambda_7$, $f^2_\infty \mu_7$ and $m$ are treated as small parameters.}
By expanding the timelike entanglement entropy \eqref{equ3.3.2} as a series \eqref{variation} in the couplings $\lambda_7$, $\mu_7$ and $m$ about (0,0,0) and substituting $t(z)$ with the complexified extremal surfaces  \eqref{RTsurface}, the leading-order gravitational corrections to holographic timelike entanglement entropy in seven-dimensional Lovelock gravity is as follows:
\[
S_A^{(T)}=&\frac{\left(\frac{1}{\epsilon^{4}}-\frac{c_6}{2}\frac{1}{(\Delta t)^{4}}\right)}{8G}+\frac{f_\infty^2 \mu_7}{4G}\left(-\frac{1}{4\epsilon^4}-\frac{3 \pi ^{5/2}  \Gamma \left(\frac{3}{5}\right)^5}{\Delta t^4 \Gamma \left(\frac{1}{10}\right)^4 \Gamma \left(\frac{21}{10}\right)}\right)+\frac{f_\infty\lambda_7}{4G}\left(\frac{1}{2\epsilon^4}-\frac{4 \pi ^{5/2} \Gamma \left(\frac{3}{5}\right)^5}{\Delta t^4 \Gamma \left(\frac{1}{10}\right)^4 \Gamma \left(\frac{11}{10}\right)}\right)\nn\\
+&\frac{\Delta t^2 m \Gamma \left(\frac{1}{10}\right)^3}{224\ 2^{4/5} \pi G \Gamma \left(-\frac{3}{10}\right) \Gamma \left(\frac{3}{5}\right)}+\frac{3 {\Delta t}^2 f_\infty^2\mu_7 m  \Gamma \left(\frac{1}{10}\right)^3}{448\ 2^{4/5} \pi G  \Gamma \left(\frac{3}{5}\right) \Gamma \left(\frac{7}{10}\right)}+\frac{7 {\Delta t}^2 f_\infty \lambda_7 m \Gamma \left(\frac{1}{10}\right) \Gamma \left(\frac{11}{10}\right)^2}{400\ 2^{4/5} \pi G  \Gamma \left(\frac{3}{5}\right) \Gamma \left(\frac{17}{10}\right)}+\dots\label{equ3.3.3}
\]
where $``\dots"$ represents the subleading contributions. Equation~\eqref{equ3.3.3} presents the timelike entanglement entropy in seven bulk dimensions, where both quadratic ($\lambda_7$) and cubic ($\mu_7$) Lovelock couplings contribute. The first line encodes the vacuum divergences: the universal UV divergence $\epsilon^{-4}$ as well as the interval-dependent contribution $(\Delta t)^{-4}$. Higher-curvature corrections enter through the $\lambda_7$ (Gauss-Bonnet) and $\mu_7$ (cubic Lovelock) terms, which modify both divergent and finite coefficients with distinct Gamma-function structures. The second line contains the leading excitation corrections, scaling as $\Delta t^2 m$, which are further modulated by the higher-curvature couplings. In particular, the coefficients of these excitation terms explicitly separate the Einstein contribution, the cubic Lovelock correction (proportional to $f_\infty^2 \mu_7$), and the Gauss-Bonnet correction (proportional to $f_\infty \lambda_7$). The appearance of different Gamma-function ratios in each sector reflects the dimension-specific analytic continuation from spacelike to timelike intervals. At the same time, the overall structure confirms the general pattern that gravitational couplings renormalize both the divergent and finite pieces of the entropy.

\subsection{Timelike entanglement entropy in $d+1$-dimensional Lovelock gravity}
Following the analysis in seven-dimensional Lovelock gravity, the discussion is extended to general $(d+1)$-dimensional Lovelock gravity, with a focus on the corresponding corrections to timelike entanglement entropy. For tractability, the Lovelock action~\eqref{equ2.2.1} is truncated at $p_{\text{max}} = 3$:
\[
I=&\frac{1}{2\ell_P^{d-1}}\int \dd^{d+1}x \sqrt{-g}\left[\frac{d(d-1)}{L^2}+R+\frac{L^2\lambda}{(d-2)(d-3)}\mathcal{L}_4-\frac{3L^4\mu}{(d-2)(d-3)(d-4)(d-5)}\mathcal{L}_6\right]+\dots
\]
The holographic timelike entanglement entropy formula for Lovelock gravity can be expressed as
\[
S_A^{(T)}=&\frac{2\pi}{\ell_p^{d-1}}\int_M\dd^{d-1}x\sqrt{h}\left[1+\frac{2L^2\lambda}{(d-2)(d-3)}\mathcal{R}-\frac{3L^4\mu}{(d-2)(d-3)(d-4)(d-5)}\mathcal{L}_4\right]\nn\\
+&\text{surface term}\label{equ3.4.1}
\]
where the surface term \cite{Myers:1987yn} can be expressed as
\[
&\text{surface term}=\frac{2\pi}{\ell_p^{d-1}}\int_{\partial M}\dd^{d-2}x\sqrt{\gamma}\left[\frac{4\lambda L^2}{(d-2)(d-3)}\mathcal{K}\right.\nn\\
-&\left.\frac{3\mu L^4}{(d-2)(d-3)(d-4)(d-5)}\left(4\mathcal{R}^B\mathcal{K}-8\mathcal{R}_{ij}^B\mathcal{K}^{ij}-\frac{4}{3}\mathcal{K}^3+4\mathcal{K}\mathcal{K}_{ij}\mathcal{K}^{ij}-\frac{8}{3}\mathcal{K}_{ij}\mathcal{K}^{jk}\mathcal{K}_k^i\right)\right]
\]
and all quantities with a subscript $B$ are evaluated on $\partial M$. 

The excited state introduced in Subsection~\ref{subsection3.2} is again considered. As noted in~\cite{Misner:1973prb}, the following expressions hold:
\[
&\mathcal{R}_{\mu\nu\rho\sigma}\mathcal{R}^{\mu\nu\rho\sigma}=\frac{d-2}{\Tilde{L}^4 \left(\left(m z^d-1\right) \Dot{t}^2+m z^d+1\right)^4}\nn\\
&\left(\Dot{t} \left(2 \left(\left(d^2+6 d-2\right) m^2 z^{2 d}-2 (d-1)\right) \Dot{t}+\left(m z^d \left(\left(d^2+6 d-2\right) m z^d-2 (d-1)\right)+2 (d-1)\right) \Dot{t}^3\right.\right.\nn\\
&+\left.4 z \left(m z^d-1\right) \Ddot{t} \left(\left((d+2) m z^d-2\right) \Dot{t}^2+(d+2) m z^d+2\right)+4 z^2 \left(m z^d-1\right)^2 \Dot{t} \Ddot{t}^2\right)\nn\\
&\left.+m z^d \left(\left(d^2+6 d-2\right) m z^6+4 (2 d-1)\right)+2 (d-2)\right);\nn\]
\[&\mathcal{R}_{\mu\nu}\mathcal{R}^{\mu\nu}=\frac{(d-2) }{\Tilde{L}^4 \left(\left(m z^d-1\right) \Dot{t}^2+m z^d+1\right)^4}\nn\\
&((d-2) (d-1)+m z^d \left(\left(\frac{d^3}{4}+\frac{11 d^2}{4}-7 d+2\right) m z^d+2 (d-1) (2 d-1)\right)+\Dot{t}\nn\\
& \left(2 z \left(m z^d-1\right) \Ddot{t} \left(\left(\frac{1}{2} \left(d^2+3 d-8\right) m z^d-2 (d-2)\right) \Dot{t}^2+\frac{1}{2} \left(d^2+3 d-8\right) m z^d+2 (d-2)\right)\right.\nn\\
&+\Dot{t} \left(\left(\frac{d^3}{2}+\frac{11 d^2}{2}-14 d+4\right) m^2 z^{2 d}+\left(m z^d \left(\left(\frac{d^3}{4}+\frac{11 d^2}{4}-7 d+2\right) m z^d-2 (d-1) (2 d-1)\right)\right.\right.\nn\\
&\left.\left.\left.+(d-2) (d-1)\right) \Dot{t}^2-2 (d-2) (d-1)\right)+(d-1) z^2 \left(m z^d-1\right)^2 \Dot{t} \Ddot{t}^2\right));\nn\\
&\mathcal{R}^B=\mathcal{R}^B_{ij}\mathcal{K}^{ij}=0;\nn\\
&\mathcal{K}\mathcal{K}_{ij}\mathcal{K}^{ij}=(d-2)^2\left(\frac{1}{\Tilde{L}^2\left(1+mz^d+(-1+mz^d)\Dot{t}^2\right)}\right)^\frac{3}{2};\nn\\
&\mathcal{K}_{ij}\mathcal{K}^{jk}\mathcal{K}_k^i=(d-2)\left(\frac{1}{\Tilde{L}^2\left(1+mz^d+(-1+mz^d)\Dot{t}^2\right)}\right)^\frac{3}{2}.\label{equ3.4.2}
\]
Thus the holographic timelike entanglement entropy in $d+1$-dimensional Lovelock gravity then becomes
\[
&S_A^{(T)}=\frac{2\pi\Tilde{L}^{d-1}}{\ell_p^{d-1}}\int_\epsilon^{z_t'}\dd z\frac{1}{z^{d-1} \left(\left(m z^d-1\right) \Dot{t}^2+m z^d+1\right)^{3/2}}\nn\\
&\left(2 \left(m z^d-1\right) \Dot{t}^2 \left(m z^d+f_\infty \lambda +1\right)+\left(m z^d+1\right) \left(m z^d+2 f_\infty \lambda +1\right)+\left(m z^d-1\right)^2 \Dot{t}^4+f_\infty^2 \mu \right),\label{equ3.4.3}
\]
where the volume of $\mathbb{R}^{d-2}$ is still normalized to unity.
{The e.o.m. derived from the functional \eqref{equ3.4.3} 
\[
\frac{\Dot{t} \left(m z^d-1\right) \left(\left(m \left(\Dot{t}^2+1\right) z^d-\Dot{t}^2+1\right) \left(-2 f_\infty \lambda +\Dot{t}^2 \left(m z^d-1\right)+m z^d+1\right)-3 f_\infty^2 \mu \right)}{z^{d-1} \left(m \left(\Dot{t}^2+1\right) z^d-\Dot{t}^2+1\right)^{5/2}}=-\frac{1}{z_t'^{d-1}}
\]admits a perturbative solution of the form\[
\Dot{t}=&(1+2f_\infty\lambda)\left(\frac{z^{2d-2}}{z^{2d-2}-z_t'^{2d-2}}\right)^\frac{1}{2}+\frac{m \left(2 z^{2d-2}-3 z_t'^{2d-2}\right) \left(\frac{z^{2d-2}}{z^{2d-2}-z_t'^{2d-2}}\right)^{3/2}}{2 z^{d-2}}\nn\\
-&3f_\infty^2\mu\frac{z^{2d-2}}{z_t'^{2d-2}}\left(\frac{z^{2d-2}}{z^{2d-2}-z_t'^{2d-2}}\right)^{-\frac{1}{2}}
\]
when $f_\infty \lambda$, $f^2_\infty \mu$ and $m$ are treated as small parameters.}
The leading-order gravitational corrections to the holographic timelike entanglement entropy in $(d+1)$-dimensional Lovelock gravity are obtained by expanding the expression~\eqref{equ3.4.3} as a series~\eqref{variation} in the couplings $\lambda$, $\mu$, and $m$ around the point $(0,0,0)$, and by substituting $t(z)$ with the complexified extremal surface~\eqref{RTsurface}:
\[
S_A^{(T)}=&\frac{\left(\frac{1}{\epsilon^{d-2}}+\frac{c_d}{2}\frac{(-i)^d}{(\Delta t)^{d-2}}\right)}{2(d-2)G}+\frac{f_\infty\lambda}{(d-2)^2G}\left(\frac{2}{ \epsilon^{d-2}}-\frac{i^{2-d} c_d}{2 \Delta t^{d-2}}\frac{\Gamma\left(\frac{d}{2(d-1)}\right)^2}{\Gamma\left(\frac{1}{2(d-1)}+1\right)\Gamma\left(\frac{1}{2(d-1)}\right)}\right)\nn\\
-&\frac{f_\infty^2\mu}{(d-2)^2G}\left(\frac{1}{\epsilon^{d-2}}+\frac{i^{2-d}3(d-1) c_d}{4(2d-1) \Delta t^{d-2}}\frac{\Gamma\left(\frac{d}{2(d-1)}\right)^2}{\Gamma\left(\frac{1}{2(d-1)}+1\right)\Gamma\left(\frac{1}{2(d-1)}\right)}\right)\nn\\
+&\frac{m \Delta t^2}{8(d-2)G}\left(\frac{1}{d+1}-\frac{2f_\infty\lambda}{d-3}-\frac{3(d-1)f^2_\infty\mu}{(d+1)(d-3)}\right)\frac{\Gamma\left(\frac{1}{2(d-1)}\right)^2\Gamma\left(\frac{1}{(d-1)}\right)}{\sqrt{\pi}\Gamma\left(\frac{1}{(d-1)}-\frac{1}{2}\right)\Gamma\left(\frac{d}{2(d-1)}\right)^2}+\dots\label{equ3.4.4}
\]
where $``\dots"$ represents the subleading contribution in $d+1$-dimensional Lovelock gravity. Equation~\eqref{equ3.4.4} shows the timelike entanglement entropy in general $(d+1)$-dimensional Lovelock gravity, where Gauss-Bonnet ($\lambda$) and cubic ($\mu$) couplings contribute on top of the Einstein term. The first line encodes the vacuum sector, containing the universal UV divergence $\epsilon^{-(d-2)}$ and a $(\Delta t)^{-(d-2)}$ contribution. Notably, the factors of $i^{2-d}$ indicate that gravitational corrections can, in general, produce an imaginary part, depending on the spacetime dimension, reflecting the analytic continuation from spacelike to timelike intervals. The second line arises from cubic Lovelock interactions, which further shift the divergent.

\subsection{Timelike entanglement entropy and entanglement entropy in Lovelock gravity}
 The literature~\cite{Guo:2025pru} demonstrates that the timelike entanglement entropy of a timelike subsystem can be entirely expressed in terms of the entanglement entropy of a corresponding spacelike subsystem, i.e.,
\[
S_{\rm bh}^{(T)}(0,0;t_0,0)&=
\frac{1}{2}\left(S_{\rm bh}(0,-t_0;0,0)+S_{\rm bh}(0,0;0,t_0)\right)\nn\\
&-\frac{d-2}{d-1}(\delta_m S(0,-t_0;0,0)+\delta_m S(0,0;0,t_0))\nn\\
&+\frac{i\left[(-i)^{d-2}-1\right]}{(d-2)\pi}\int_{-t_0}^{t_0}dxx^{d-2}\partial_tS_{\rm bh}(0,x;0,0)\,,
\label{eq:duality-time-space}
\]
where $S_{\rm bh}(a,b;c,d)$ represents the entanglement entropy for a boundary subsystem with endpoints at $(t_1,x_1)=(a,b)$ and $(t_2,x_2)=(c,d)$, and $\delta_m S$ denotes the first-order correction to the entanglement entropy due to the black hole mass. In the black hole background, the entanglement entropies appearing on the right-hand side of the relation correspond to spacelike subsystems, and their associated Ryu–Takayanagi (RT) surfaces remain entirely outside the event horizon. This correspondence suggests that timelike entanglement entropy can serve as a probe of geometric information behind the horizon. The following analysis investigates whether this relation persists in the presence of higher-curvature gravitational corrections. Specifically, gravitational corrections to spacelike entanglement entropy are first computed for strip-shaped subsystems in $(d+1)$-dimensional Lovelock gravity, followed by a systematic comparison with the timelike case.

The spacelike strip is defined as
\[
A=\left\{(t,\mathbf{x}): t=0, x_1\in \left[-\frac{a}{2},\frac{a}{2}\right], \mathbf{x}_{\|}\in\mathbb{R}^{d-2}\right\}.
\]
In $d>2$, the holographic entanglement entropy in the vacuum is known \cite{Doi:2023zaf}:
\[
S_A=\frac{\left(\frac{1}{\epsilon^{d-2}}+\frac{c_d}{2}\frac{1}{a^{d-2}}\right)}{2(d-2)G},\quad c_d=\left(\frac{2\sqrt{\pi}\Gamma\left(\frac{d}{2(d-1)}\right)}{\Gamma\left(\frac{1}{2(d-1)}\right)}\right)^{d-1},\label{EE}
\]
and the extremal surface $\gamma_{A}$ takes the form
\[
\mathbf{X}^\mu=\left\{t=0,x{_{\pm}(z)},z, \mathbf{x}_{\|}, x_\perp=0\right\},
\]
where
\[
x_\pm(z)=&\pm\left(\frac{a}{2}-\frac{z_* \left(\frac{z}{z_*}\right)^d \, _2F_1\left(\frac{1}{2},\frac{d}{2 (d-1)};\frac{3 d-2}{2 (d-1)};\left(\frac{z}{z_*}\right)^{2 d-2}\right)}{d}\right) \ \text{with}\ z_*=\frac{a \Gamma \left(\frac{1}{2 (d-1)}\right)}{2 \left(\sqrt{\pi } \Gamma \left(\frac{d}{2 (d-1)}\right)\right)}.\label{spacelikeRT}
\]
The holographic  entanglement entropy formula for Lovelock gravity can be expressed as
\[
S_{A}=\frac{2\pi\Tilde{L}^{d-1}}{\ell_p^{d-1}}\int_\epsilon^{z_*}\dd z\frac{2 \Dot{x}^2 \left(m z^d+f_\infty \lambda +1\right)+\left(m z^d+1\right) \left(m z^d+2 f_\infty \lambda+1\right)+f_\infty^2 \mu+\Dot{x}^4}{z^{d-1} \left(m z^d+\Dot{x}^2+1\right)^{3/2}}.\label{equ3.5.1}
\]
{The e.o.m. derived from the functional \eqref{equ3.5.1} 
\[\frac{\Dot{x} \left(-2 f_\infty {\lambda} \left(m z^d+\Dot{x}^2+1\right)+\left(m z^d+\Dot{x}^2+1\right)^2-3 f^2_\infty {\mu }\right)}{\left(m z^d+\Dot{x}^2+1\right)^{5/2}}
=\frac{1}{z_*'^{d-1}}
\]admits a perturbative solution of the form\[
\Dot{x}=&(1+2f_\infty\lambda+\frac{1}{2}mz^d)\left(\frac{z^{2d-2}}{z_*'^{2d-2}-z^{2d-2}}\right)^\frac{1}{2}+3f_\infty^2\mu\frac{z^{2d-2}}{z_*'^{2d-2}}\left(\frac{z^{2d-2}}{z_*'^{2d-2}-z^{2d-2}}\right)^{-\frac{1}{2}}
\]
when $f_\infty \lambda_7$, $f^2_\infty \mu_7$ and $m$ are treated as a small parameters.}
The leading-order gravitational corrections to the holographic entanglement entropy in $(d+1)$-dimensional Lovelock gravity are obtained by expanding the entanglement entropy expression~\eqref{equ3.5.1} as a series~\eqref{variation} in the couplings $\lambda$, $\mu$, and $m$ around the point $(0,0,0)$, and by substituting $x(z)$ with the complexified extremal surface~\eqref{spacelikeRT}:
\[
S_A=&\frac{\left(\frac{1}{\epsilon^{d-2}}+\frac{c_d}{2}\frac{1}{a^{d-2}}\right)}{2(d-2)G}+\frac{f_\infty\lambda}{(d-2)^2G}\left(\frac{2}{ \epsilon^{d-2}}-\frac{ c_d}{2 a^{d-2}}\frac{\Gamma\left(\frac{d}{2(d-1)}\right)^2}{\Gamma\left(\frac{1}{2(d-1)}+1\right)\Gamma\left(\frac{1}{2(d-1)}\right)}\right)\nn\\
-&\frac{f_\infty^2\mu}{(d-2)^2G}\left(\frac{1}{\epsilon^{d-2}}+\frac{3(d-1) c_d}{4(2d-1) a^{d-2}}\frac{\Gamma\left(\frac{d}{2(d-1)}\right)^2}{\Gamma\left(\frac{1}{2(d-1)}+1\right)\Gamma\left(\frac{1}{2(d-1)}\right)}\right)\nn\\
+&\frac{m  a^2}{8(d-2)G}\left(\frac{d-1}{(d+1)(3-d)}-\frac{2f_\infty\lambda(d-1)}{(d-3)(d+1)}-\frac{9(d-1)^2f^2_\infty\mu}{(d+1)(3-d)(3d-1)}\right)\frac{\Gamma\left(\frac{1}{2(d-1)}\right)^2\Gamma\left(\frac{1}{(d-1)}\right)}{\sqrt{\pi}\Gamma\left(\frac{1}{(d-1)}-\frac{1}{2}\right)\Gamma\left(\frac{d}{2(d-1)}\right)^2}\nn\\
+&\dots\label{spacelikeresult}
\]
Let each term in the timelike entanglement entropy in Lovelock gravity be denoted by $\alpha_{m^i\lambda^j\mu^k}$, where the indices $i, j, k \in \{0,1\}$ indicate the order of contributions from excited states ($m$), and gravitational couplings ($\lambda$ and $\mu$). For example, a term of order $m\mu$ is written as $\alpha_{m^1\lambda^0\mu^1}$. Similarly, the corresponding terms in the spacelike entanglement entropy are denoted as $\beta_{m^i\lambda^j\mu^k}$. By identifying $a = \Delta t$, the following relations are obtained:
\[
&\frac{\alpha_{m^0\lambda^1\mu^0}}{\beta_{m^0\lambda^1\mu^0}}=\frac{\alpha_{m^0\lambda^0\mu^1}}{\beta_{m^0\lambda^0\mu^1}}=-(i)^{-d},\quad\frac{\alpha_{m^1\lambda^0\mu^0}}{\beta_{m^1\lambda^0\mu^0}}=-\frac{d-3}{d-1},\nn\\
&\frac{\alpha_{m^1\lambda^1\mu^0}}{\beta_{m^1\lambda^1\mu^0}}=-\frac{ d+1}{ d-1},\quad\frac{\alpha_{m^1\lambda^0\mu^1}}{\beta_{m^1\lambda^0\mu^1}}=-\frac{3 d-1}{3 d-3}.\label{alpha_beta}
\]
The comparison shows that the corresponding coefficients follow fixed proportionalities between the timelike \eqref{equ3.4.4} and spacelike \eqref{spacelikeresult} cases. 
For the vacuum curvature corrections, the mapping is governed by a universal analytic-continuation phase, with $a \to i \Delta t$ introducing relative factors such as $i^{2-d}$. 
For the excitation sector, the coefficients in the Einstein, Gauss-Bonnet, and cubic Lovelock parts are linked by simple rational prefactors that depend only on the spacetime dimension $d$. It should also be emphasized that in the absence of excitations, the perturbative result, namely the AdS gravitational vacuum correction, is obtained simply by analytically continuing $a \to i \Delta t$. 
This suggests that even beyond perturbation theory, the essential difference between spacelike and timelike intervals may still be captured by such a straightforward analytic continuation of the subsystem.  

Taken together, these results demonstrate that timelike and spacelike entanglement entropies are not independent but are connected by precise relations: the vacuum terms differ by analytic-continuation phases, whereas the excitation terms are related by dimension-dependent rational ratios \eqref{alpha_beta}. This provides a clear and systematic map between the two cases, disentangling the respective roles of vacuum geometry and low-energy excitations.

The relation in Eq.~\eqref{alpha_beta} shows that each term in both timelike and spacelike entanglement entropy can be expressed by a universal factor depending only on the dimension $d$. Moreover, due to the presence of the factors $(\Delta x^2-\Delta t^2)^\frac{1-d}{2}$ and $(\Delta x^2-\Delta t^2)^\frac{2-d}{2}$ in the $\lambda$ and $\mu$ correction terms in Eq.~\eqref{spacelikeresult}\footnote{The derivative of $\delta S$ with respect to $t$ is evaluated by invoking Lorentz symmetry in the $\{t, \vec{x}\}$ directions and expressing $a^2$ as $a^2 \equiv \Delta x^2 - \Delta t^2$.}, the computation of $\partial_{\Delta t} \delta S(\Delta t,\Delta x,0,0)|_{\Delta t=0}$ yields zero. As a result, we can replace $S_{\rm Ein}$ in the integral of Eq.~\eqref{eq:duality-time-space} with $S_{\rm Ein} + \delta_\lambda S + \delta_\mu S$. Finally, the timelike entanglement entropy including higher-order gravitational corrections can be written in a form similar to Eq.~\eqref{eq:duality-time-space}, i.e.,
\[
S_{\rm LL}^{(T)}(0,0;t_0,0)&=
\frac{1}{2}\left(S_{\rm LL}(0,-t_0;0,0)+S_{\rm LL}(0,0;0,t_0)\right)\nn\\
&-\frac{d-2}{d-1}(\delta_m S(0,-t_0;0,0)+\delta_m S(0,0;0,t_0))\nn\\
&-\frac{(-i)^d+1}{2}(\delta_\lambda S(0,-t_0;0,0)+\delta_\lambda S(0,0;0,t_0))\nn\\
&-\frac{(-i)^d+1}{2}(\delta_\mu S(0,-t_0;0,0)+\delta_\mu S(0,0;0,t_0))\nn\\
&-\frac{d}{d-1}(\delta_{m,\lambda} S(0,-t_0;0,0)+\delta_{m,\lambda} S(0,0;0,t_0))\nn\\
&-\frac{3d-2}{3d-3}(\delta_{m,\mu} S(0,-t_0;0,0)+\delta_{m,\mu} S(0,0;0,t_0))\nn\\
&+\frac{i\left[(-i)^{d-2}-1\right]}{(d-2)\pi}\int_{-t_0}^{t_0}dxx^{d-2}\partial_tS_{\rm LL}(0,x;0,0)\,.
\]
Although this expression is not particularly compact, it nonetheless shows that timelike entanglement entropy can still be expressed entirely in terms of spacelike entanglement entropy.


\section{Timelike entanglement entropy for hyperbolic subsystem in Lovelock gravity}\label{section4}
Following the analysis of holographic timelike entanglement entropy for strip-shaped subsystems, this section turns to the case of hyperbolic subsystems in the context of higher-derivative gravitational theories.\footnote{See also the recent discussion in~\cite{Nunez:2025puk} regarding timelike entanglement entropy for hyperbolic subsystems.} The hyperbolic subsystem under consideration is defined as
\[
A = \left\{ (t, x_1, \dots, x_{d-1}) \;\middle|\; t^2 - x_1^2 - \dots - x_{d-2}^2 \leq R^2,\; x_{d-1} = 0 \right\}.
\]
The vacuum contribution to the holographic timelike entanglement entropy is computed using complex extremal surfaces. Higher-order gravitational corrections are then evaluated within the framework of Lovelock gravity.
\subsection{Timelike entanglement entropy for hyperbolic subsystem in the vacuum}
Given the boundary's SO(1,d-2) symmetry, the extremal surface must inherit this symmetry. Using hyperbolic coordinates, it can be parametrized as
\[
&t=\rho(z) \cosh(\psi),\nn\\
&x_i=\rho(z)\sinh(\psi)\cdot\hat{n}_i,\quad\left(\sum\limits_{i=1}^{n-3}\hat{n}_i^2=1\right),
\]
where $z$ is a  complex coordinate and $\rho(z)$ is a complex function.
The induced metric on the extremal surface $\gamma_A$ derived from the $AdS_{d+1}$ metric is 
\[
ds_{induced}^2=\frac{L^2}{z^2}\left[\left(1-\Dot{\rho}^2\right)dz^2+\rho^2d\psi^2+\rho^2\sinh^2\psi d\Omega_{d-3}^2\right].\label{hyperboli_metric}
\]
The function $\rho(z)$ is found by minimizing the area functional
\[
\mathcal{A}_{\gamma_A}=L^{d-1}Vol(\mathbb{H}^{d-2})\int \dd z\frac{\rho^{d-2}}{z^{d-1}}\sqrt{1-\Dot{\rho}^2}.\label{equ4.1}
\]
The e.o.m. of $\eqref{equ4.1}$ is
\[
(d-2)z(1-\Dot{\rho}^2)+\rho\Ddot{\rho}z(1-\Dot{\rho}^2)+(1-d)\rho\Dot{\rho}(1-\Dot{\rho}^2)+\rho\Dot{\rho}^2z\Ddot{\rho}=0
\]
and has the following simple solution
\[
-z^2+\rho^2=R^2.\label{RT_hyperbolic}
\]
While the functional form of the extremal surface matches that given in~\cite{Doi:2023zaf}, the present construction is formulated as a curve embedded in the complexified space $\mathbb{C}
$, rather than in a real manifold. Using the expression~\eqref{RT_hyperbolic}, the holographic timelike entanglement entropy for the hyperbolic subsystem in the vacuum is obtained as follows:
\[
&S_A^{(T)}=\frac{L^{d-1}}{4G_{N}^{d+1}}Vol(\mathbb{H}^{d-2})\int_\epsilon^{z_\rho} \dd z\frac{\rho^{d-2}}{z^{d-1}}\sqrt{1-\Dot{\rho}^2}\nn\\
&=\frac{L^{d-1}}{4G_{N}^{d+1}}Vol(\mathbb{H}^{d-2})\begin{cases}
        \sum\limits_{k=0}^{\frac{d-3}{2}}\binom{\frac{d-3}{2}}{k}\frac{1}{d-2k-2}\left(\frac{R}{\epsilon}\right)^{d-2k-2}+\frac{i\sqrt{\pi}\Gamma\left(\frac{d-1}{2}\right)}{2\Gamma\left(\frac{d}{2}\right)},\quad (d:\ odd)\\
    \sum\limits_{k=0}^{\frac{d-4}{2}}\binom{\frac{d-3}{2}}{k}\frac{1}{d-2k-2}\left(\frac{R}{\epsilon}\right)^{d-2k-2}+\frac{\Gamma\left(\frac{d-1}{2}\right)}{\sqrt{\pi}\Gamma\left(\frac{d}{2}\right)}\log\frac{R}{2\epsilon}+\frac{i\sqrt{\pi}\Gamma\left(\frac{d-1}{2}\right)}{2\Gamma\left(\frac{d}{2}\right)},\quad (d:\ even),
\end{cases}\label{equ4.2}
\]
where $z_\rho=i R$. In contrast to the method of \cite{Doi:2023zaf}, which requires evaluating two separate integrals for the real and imaginary parts of holographic timelike entanglement entropy, or carrying out an analytic continuation only at the final step--our formalism achieves a genuinely unified description. By embedding the computation directly into the complexified space $\mathbb{C}^2$, both the real and imaginary contributions are naturally incorporated within a single integral representation. This unified framework not only streamlines the calculation but also highlights a clear geometric interpretation: the analytic continuation is no longer an external prescription but an intrinsic feature of the setup itself. As a result, our approach provides both conceptual clarity and technical efficiency in the treatment of timelike entanglement entropy.

\subsection{Timelike entanglement entropy for hyperbolic subsystem in Gauss-Bonnet gravity}
This subsection considers the computation of timelike entanglement entropy for hyperbolic subsystems in the simplest case of Lovelock gravity, namely five-dimensional Gauss-Bonnet theory. The holographic timelike entanglement entropy \eqref{equ3.1.4} evaluated with the induced metric \eqref{hyperboli_metric}\footnote{Excited states are intentionally excluded here, as mass terms break the diagonal structure of the induced metric on the extremal surface.} is
\[
S_A^{(T)}=\frac{2\pi \Tilde{L}^3}{\ell_p^3}Vol(\mathbb{H}^2)\int_{\epsilon}^{z_\rho}\dd z\frac{2 f_\infty \lambda _5 \left(z^2 \left(2 \Dot{\rho}^2-1\right)-2 z \rho \Dot{\rho}+\rho^2\right)+\rho^2 \left(1-\Dot{\rho}^2\right)}{z^3 \left(1-\Dot{\rho}^2\right)^{1/2}}.\label{equ4.2.1}
\]
It should be emphasized that the holographic timelike entanglement entropy is computed via integration over a complexified extremal surface, rather than a real submanifold.

By performing a series expansion of the entanglement entropy \eqref{equ4.2.1} in the couplings $\lambda$ about $0$, and inserting $\rho(z)$ from the complexified extremal surfaces \eqref{RT_hyperbolic}, the leading–order gravitational corrections to holographic entanglement entropy in the hyperbolic subsystem of five–dimensional Gauss–Bonnet gravity are obtained:
\[
\Delta S_A^{(T)}=\frac{L^{3}}{4G_{N}^{4}}Vol(\mathbb{H}^{2})2f_\infty\lambda_5\left(\frac{R^2}{2 \epsilon ^2}-\frac{1}{4}  \left(6\log \left(\frac{2R}{\epsilon}\right)-1 \right)-\frac{3\pi i}{4}\right)+\dots\label{equ4.2.2}
\]
where $\dots$ represents the subleading contribution in 5-dimensional Gauss-Bonnet gravity.

The calculation is next extended to Gauss–Bonnet gravity in arbitrary $(d+1)$ dimensions, yielding a corresponding expression for the timelike entanglement entropy in analogy with the previous case:
\[
S_A^{(T)}=\frac{2\pi \Tilde{L}^{d-1}}{\ell_p^{d-1}}Vol(\mathbb{H}^{d-2})\int_{\epsilon}^{z_\rho}\dd z\frac{\rho^{d-4}\left[2 f_\infty \lambda _5 \left(z^2 \left(2 \Dot{\rho}^2-1\right)-2 z \rho \Dot{\rho}+\rho^2\right)+\rho^2 \left(1-\Dot{\rho}^2\right)\right]}{z^{d-1} \left(1-\Dot{\rho}^2\right)^{1/2}}.\label{equ4.2.3}
\]
Expanding the entanglement entropy expression~\eqref{equ4.2.3} in a power series of the coupling $\lambda$ around zero and substituting $\rho(z)$ with the complexified extremal surface~\eqref{RT_hyperbolic} yields the leading-order gravitational corrections to the holographic entanglement entropy for a hyperbolic subsystem in $(d+1)$-dimensional Gauss–Bonnet gravity:
\[
\Delta S_A^{(T)}=&\frac{L^{d-1}}{4G_{N}^{d+1}}Vol(\mathbb{H}^{d-2})2f_\infty\lambda\nn\\
&\begin{cases}
      \left(\frac{1}{(d-2)\epsilon^{d-2}}-i\frac{\sqrt{\pi } (d-1)  \Gamma \left(\frac{d-3}{2}\right)}{4 \Gamma \left(\frac{d}{2}\right)}\right)  +\dots,\quad (d:\ odd)\\
  \left(\frac{1}{(d-2)\epsilon^{d-2}}+\frac{(d-1) \Gamma \left(\frac{d-3}{2}\right)}{2 \sqrt{\pi } \Gamma \left(\frac{d}{2}\right)}\log\left(\frac{\epsilon}{2R}\right)-i\frac{\sqrt{\pi } (d-1)  \Gamma \left(\frac{d-3}{2}\right)}{4 \Gamma \left(\frac{d}{2}\right)}\right)+\dots,\quad (d:\ even),
\end{cases}\label{equ4.2.4}
\]
Here, constant terms and subleading contributions in~\eqref{equ4.2.3} have been omitted. In contrast to the strip geometry discussed in Section~\ref{section3}, where higher-curvature corrections to the imaginary part of the holographic timelike entanglement entropy arise only in odd-dimensional boundary theories, the hyperbolic case exhibits a distinct qualitative behavior. In this setting, imaginary contributions appear in all dimensions, indicating that the analytic continuation affects hyperbolic slices differently and induces nonvanishing phase factors even in even-dimensional spacetimes. This feature highlights the dependence of timelike entanglement entropy on the geometry of the entangling surface, emphasizing the influence of subsystem shape on the analytic structure of higher-curvature corrections.\footnote{The analysis may, in principle, be extended to Lovelock theories in arbitrary dimensions. However, the increasing complexity introduced by higher-order curvature terms renders such computations significantly more involved. For this reason, the full set of results is not presented here.}
\section{Summary and discussion}\label{section5}

This paper has investigated holographic timelike entanglement entropy in higher-curvature gravity theories, with a particular focus on excitation states, which encode richer physical information than the vacuum.  

Starting with five-dimensional Gauss-Bonnet gravity as the simplest higher-curvature model, we computed the corresponding corrections to timelike entanglement entropy (eq. \eqref{result3.1}) and then generalized the analysis to arbitrary spacetime dimensions, thereby identifying universal correction patterns (eq. \eqref{result3.2}). We subsequently examined the minimal Lovelock theory incorporating cubic curvature interactions \eqref{equ3.3.3}, and finally extended the computation to the most general cubic Lovelock gravity in arbitrary dimensions \eqref{equ3.4.4}. Across these cases, we observed that higher-curvature corrections may modify the imaginary part of timelike entanglement entropy in a dimension-dependent way. In contrast, excitation states contribute solely to its real part.  

Since timelike entanglement entropy originates from the analytic continuation of its spacelike counterpart, we examined in detail how higher-curvature corrections transform under this continuation. The comparison shows that the coefficients in the two cases obey fixed proportionality relations: for the vacuum sector, the mapping is controlled by a universal analytic-continuation phase, with $\Delta x \to i \Delta t$ generating factors such as $i^{2-d}$; while the divergent structures remain identical, the finite terms differ through these phase factors together with dimension-specific Gamma-function ratios. In the excitation sector, by contrast, the Einstein, Gauss-Bonnet, and cubic Lovelock contributions are related by simple rational prefactors that depend only on the spacetime dimension $d$. It is worth emphasizing that in the absence of excitations, the perturbative vacuum correction in AdS gravity is obtained simply by analytically continuing $a \to i \Delta t$\footnote{Our analysis has relied on the analytic continuation to relate 
spacelike and timelike entanglement entropies. While this procedure captures the correct 
analytic structure, it does not by itself encode all causal features of the bulk geometry. 
}. This observation suggests that even beyond perturbation theory, the essential distinction between spacelike and timelike intervals may still reduce to a straightforward analytic continuation of the subsystem.  

For hyperbolic subsystems, we have shown that in the vacuum, timelike entanglement entropy can be obtained either through analytic continuation or by evaluating a complexified extremal surface \eqref{equ4.2}, with both methods yielding identical results. This consistency validates the geometric picture and, within five-dimensional and general $(d+1)$-dimensional Gauss-Bonnet gravity, we further computed higher-curvature corrections, thereby extending the use of the complex surface framework to higher-curvature corrections \eqref{equ4.2.2}. Unlike the strip case discussed in Sec.~\ref{section3}, where higher-curvature corrections to the imaginary part of timelike entanglement entropy occur only in odd-dimensional boundary theories, the hyperbolic geometry exhibits imaginary contributions in all dimensions. This highlights the universality of hyperbolic subsystems and shows that the analytic structure of timelike entanglement entropy is highly sensitive to the geometry of the entangling region.  

{
As in ordinary holographic entanglement entropy, timelike entanglement entropy exhibits
UV divergences regulated by a cutoff $\epsilon$\footnote{We are grateful to the anonymous referee for bringing this important perspective to our attention.}. We can adopt a minimal subtraction 
scheme, introducing the same geometric counterterms as in Einstein gravity but with 
coefficients modified by the higher-curvature couplings. In this way all 
power-law divergences are cancelled, while the finite remainder acquires higher-curvature 
corrections. Beyond the minimal subtraction scheme, one may also 
consider more general renormalization prescriptions\footnote{For more detailed discussions of renormalization in the context of entanglement entropy, 
we refer the reader to \cite{Taylor:2016aoi} for a general analysis and to 
\cite{Anastasiou:2021jcv,Kofinas:2007ns} for the explicit treatment in Lovelock gravity.}. The divergence structure in the 
Einstein vacuum takes the form $\sim 1/\epsilon^{\,d-2}$, while in the presence of 
higher-curvature couplings the divergences are modified only by coupling-dependent 
coefficients, schematically $\sim f(\lambda)/\epsilon^{\,d-2}$. Thus the same set of 
geometric counterterms as in the Einstein case suffices to cancel the divergences, 
with coefficients appropriately adjusted. Before renormalization, the finite part of timelike entanglement entropy can be 
schematically expressed as $S^{(T)} \sim \text{const}_1 + f(\lambda )\text{const}_2$. 
Upon renormalization, both $\text{const}_1$ and $\text{const}_2$ are subject to the 
same finite subtraction. Unless one fine-tunes 
the counterterms to eliminate higher-curvature effects altogether, the universal finite 
part of the Einstein vacuum is generically shifted by higher-curvature corrections. 
It is natural to regard the vacuum divergence of timelike entanglement entropy as a 
vacuum effect, analogous to Casimir energy. In practice, one is often interested in 
physical quantities relative to the vacuum, so different renormalization prescriptions 
for the vacuum timelike entanglement entropy correspond to different renormalization conditions on this vacuum 
piece. The higher-curvature dependence of the finite contribution, however, is a robust 
feature independent of the particular scheme.
}

The present analysis {based on Eq.~\eqref{equ3.4.1}} is carried out within the framework of Lovelock gravity, which serves as a tractable model here. {In more general higher-curvature theories, to calcuate the exact form of timelike entanglement entropy requires the use of the formula in~\cite{Dong:2013qoa}.} Our computations capture only the leading-order perturbative corrections, whereas a fully nonperturbative determination would require solving higher-order nonlinear differential or algebraic equations, many of which (such as quintic equations) cannot be expressed in terms of radicals. {In such cases one may resort to numerical approaches~\footnote{If numerically solving the minimal surface problem, one can use the shooting method, which involves assuming the value of $z_t$ at $x=\delta$, and then computing the integral curve $t(z)$ using $z_t$ and $t'(z)$. By enumerating $z_t$ within an appropriate range, the endpoint of the integral curve can be made to satisfy $x=R$ at $z=\epsilon$.} or suitable approximation schemes to explore the behavior of timelike entanglement entropy. We leave a systematic investigation of such nonperturbative regimes for future work.} Finally, in the hyperbolic case, we encountered multiple complex extremal surfaces \eqref{RT_hyperbolic}. {Unlike the case of spacelike entanglement entropy where, for instance, the RT surface
for a spherical region can be unambiguously chosen as the $z>0$ branch, in the timelike
setup the extremal surfaces live in a complexified bulk and such a direct prescription
is absent.} In this work, we selected the saddle that reproduces the analytic continuation result, but the question of how to systematically identify the physically relevant surface remains unsettled. Proposals in \cite{Heller:2024whi, Heller:2025kvp} suggest criteria such as choosing the surface with the smallest real part or one consistent with analytic continuation. In addition to the aforementioned solutions, the authors in~\cite{Li:2022tsv} propose a new definition of extremal surfaces, namely the Complex-valued Weak Extremal Surface, to address the multivaluedness issue of timelike entanglement. However, a general principle for selecting among competing saddles is still lacking. Clarifying this issue will be an important direction for future research.

\appendix

\acknowledgments

We thank Wu-zhong Guo, Yun-Ze Li, Hao Ouyang, Yuan Sun and Yu-Xuan Zhang for valuable discussions on this work.
L. Z. and Z. Z. are supported by the Science and Technology Development Plan Project of Jilin Province, China Grant No. 20240101326JC.
S. H. acknowledges financial support from the Max Planck Partner Group and the Natural Science Foundation of China, Grants No. 12475053 and No. 12235016.


 \bibliographystyle{JHEP}
 \bibliography{biblio.bib}


\end{document}